\documentclass{iopart}
\usepackage{iopams}

\begin{document}
\title[Saari's homographic conjecture]{Saari's homographic conjecture
for planar equal-mass three-body problem
in Newton gravity}

\author{
Toshiaki Fujiwara$^1$,
Hiroshi Fukuda$^2$,
Hiroshi Ozaki$^3$\\
and Tetsuya Taniguchi$^4$
}

\address{
$^{1}$ $^{2}$ $^{4}$College of Liberal Arts and Sciences, Kitasato University,
1-15-1 Kitasato, Sagamihara, Kanagawa 252-0329, Japan
}

\address{
$^3$General Education Program Center, Tokai University, Shimizu Campus,
3-20-1, Orido, Shimizu, Shizuoka 424-8610, Japan
}

\eads{
$^{1}$fujiwara@kitasato-u.ac.jp,
$^{2}$fukuda@kitasato-u.ac.jp,
$^{3}$ozaki@tokai-u.jp,
$^{4}$tetsuya@kitasato-u.ac.jp
}

\begin{abstract}
Saari's homographic conjecture 
in $N$-body problem under the Newton gravity is the following;
configurational measure $\mu=\sqrt{I}\, U$,
which is the product of square root of the moment of inertia
$I=(\sum m_k)^{-1}\sum m_i m_j r_{ij}^2$
and the potential function $U=\sum m_i m_j/r_{ij}$,
is constant if and only if
the motion is homographic.
Where $m_k$ represents mass of body $k$
and $r_{ij}$ represents distance between bodies $i$ and $j$.
We prove this conjecture for planar equal-mass 
three-body problem.

In this work, we use three sets of shape variables.
In the first step, we use $\zeta=3q_3/(2(q_2-q_1))$
where $q_k \in \mathbb{C}$ represents position of body $k$.
Using 
$r_1=r_{23}/r_{12}$ and $r_2=r_{31}/r_{12}$
in intermediate step,
we finally use
$\mu$ itself and $\rho=I^{3/2}/(r_{12}r_{23}r_{31})$.
The shape variables $\mu$ and $\rho$
make our proof simple.
\end{abstract}

\submitto{\JPA}

\maketitle

\section{Saari's homographic conjecture}
In 2005, Donald Saari formulated his conjecture
in the following form \cite{SaariCollisions,SaariSaarifest};
in the $N$-body problem under
the potential function
\begin{equation}
\label{eq:U}
U=\sum_{1\le i<j\le N} \frac{m_i m_j}{r_{ij}^\alpha},
\quad \alpha>0,
\end{equation}
a motion has a constant configurational measure
\begin{equation}
\label{eq:mu}
\mu = I^{\alpha/2}\, U
\end{equation}
if and only if the motion is homographic.
Here, 
$r_{ij}$ represents the mutual distance
between the bodies $i$ and $j$,
and $I$ represents the moment of inertia
\begin{equation}
\label{eq:I}
I=(\sum_{1\le k \le N} m_k)^{-1} \sum_{1\le i<j\le N} m_i m_j r_{ij}^2.
\end{equation}

Florin Diacu, Toshiaki Fujiwara, Ernesto Perez-Chavela
and Manuele Santoprete called this conjecture
the ``Saari's homographic conjecture''
and partly proved this conjecture for some cases \cite{DiacHomographic}.
Recently,
the present authors proved this conjecture
for planar equal-mass  three-body problem for $\alpha=2$ \cite{Fujiwara2011}.
In this paper,
we extends our proof to $\alpha=1$,
the Newton gravity.

In section \ref{sec:eqOfMotion},
we derive the equations of motion for the size change, rotation
and shape change.
To do this, we use the shape variable $\zeta$,
\begin{equation}
\label{eq:zeta}
\zeta = \frac{3}{2}\frac{q_3}{q_2-q_1},
\end{equation}
introduced by Richard Moeckel and Richard Montgomery \cite{M&M}.
Here, $q_k \in \mathbb{C}$, $k=1,2,3$ represents
position of the body $k$.
Then, in the section \ref{sec:nesCond},
we investigate motions with $\mu=$ constant
and non-homographic,
and we derive a necessary condition
that must be satisfied by such motion.
The contents in the sections \ref{sec:eqOfMotion} and \ref{sec:nesCond} are
review of our previous paper \cite{Fujiwara2011},
although we  changed few notations. 
To prove the Saari's conjecture,
we will show that no finite orbit satisfies the necessary condition.
To attain this purpose, the expression of the necessary condition
by $\zeta$ is too complex.
To simplify the expression,
we will use other set of shape variables,
\begin{equation}
\label{def:r1andr2}
r_1=|\zeta-1/2|=r_{23}/r_{12},\quad
r_2=|\zeta+1/2|=r_{31}/r_{12}.
\end{equation}
Then, using the invariance of the system under the 
permutations of $\{q_1, q_2, q_3\}$,
we rewrite the necessary condition in  another set of shape variables
$\mu$ itself and $\rho$,
\begin{equation}
\mu=I^{1/2}\left(\frac{1}{r_{12}}+\frac{1}{r_{23}}+\frac{1}{r_{31}}\right),
\quad
\rho=\frac{I^{3/2}}{r_{12}r_{23}r_{31}},
\end{equation}
that are manifestly invariant under the permutations.
Since, we are considering $\mu=$ constant orbits,
variables $\mu$ and $\rho$ make our proof  easy.
This expression is given in section \ref{sec:invariance}.
The proof of the Saari's conjecture is given in the section \ref{sec:proof}.
In section \ref{sec:discussions}, we give discussions.

\section{Equations of motion}
\label{sec:eqOfMotion}
In this section,
we summarize the equations of motion for $\alpha=1$
in terms of size, rotation and shape.
We don't assume $\mu=$ constant in this section.

Let $q_k \in \mathbb{C}$ be the position 
and mass $m_k=1$ for $k=1,2,3$.
We take the center of mass frame,
$\sum q_k=0$.
The Lagrangian is given by,
\begin{equation}
L=\frac{1}{2}\sum \left| \frac{dq_k}{dt} \right|^2 +U.
\end{equation}
%

We take the shape variable $\zeta \in \mathbb{C}$
in (\ref{eq:zeta}).
This variable is invariant under the scaling and rotation,
$q_k \to \lambda e^{i\theta}q_k$ with $\lambda, \theta \in \mathbb{R}$.
Thus, $\zeta$ depends only on shape.
Let us define $\xi_k=q_k/(q_2-q_1)$. Then, we have,
\begin{equation}
\label{eq:xi}
\xi_1=-\frac{1}{2}-\frac{\zeta}{3},\quad
\xi_2=\frac{1}{2}-\frac{\zeta}{3},\quad
\xi_3=\frac{2\zeta}{3}.
\end{equation}
Since, the triangle $q_1 q_2 q_3$ and $\xi_1 \xi_2 \xi_3$
are similar and have the same orientation,
we have two variables $I \ge 0$ and $\theta \in \mathbb{R}$,
such that
\begin{equation}
q_k = \sqrt{I}\, e^{i\theta} \frac{\xi_k}{\sqrt{\sum |\xi_\ell|^2}}.
\end{equation}
We take $I$, $\theta$ and $\zeta$ for dynamical variables.
In the following, we identify
$\zeta=x+iy$ and $\mathbf{x}=(x,y) \in \mathbb{R}^2$.
By direct calculations, we obtain the Lagrangian
\begin{equation}
L
=\frac{\dot{I}^2}{8I}
	+\frac{I}{2}
		\left(
			\dot{\theta}
			+\frac{\frac{4}{3}\mathbf{x} \wedge \dot{\mathbf{x}}}
				{1+\frac{4}{3}|\mathbf{x}|^2}
		\right)^2
	+\frac{I}{2}
		\frac{\frac{4}{3}|\dot{\mathbf{x}}|^2}
			{(1+\frac{4}{3}|\mathbf{x}|^2)^2}
	+\frac{\mu(\mathbf{x})}{\sqrt{I}}.
\end{equation}
Here, $\,\dot{}\,$ represents time derivative,
$\mathbf{x} \wedge \dot{\mathbf{x}}=x\dot{y}-y\dot{x}$
and
\begin{equation}
\mu(\mathbf{x})
=\sqrt{\frac{1}{2}+\frac{2}{3}|\mathbf{x}|^2}
	\left(
		1
		+\frac{1}{\sqrt{(x-1/2)^2+y^2}}
		+\frac{1}{\sqrt{(x+1/2)^2+y^2}}
	\right).
\end{equation}

Since, $\theta$ is cyclic, the angular momentum $C$ is constant
of motion,
\begin{equation}
C
=\frac{\partial L}{\partial \dot{\theta}}
=I \left(
			\dot{\theta}
			+\frac{\frac{4}{3}\mathbf{x} \wedge \dot{\mathbf{x}}}
				{1+\frac{4}{3}|\mathbf{x}|^2}
	\right).
\end{equation}
Therefore, the total energy $E$ is given by
\begin{equation}
E=\frac{\dot{I}^2}{8I}
	+\frac{C^2}{2I}
	+\frac{I}{2}
		\frac{\frac{4}{3}|\dot{\mathbf{x}}|^2}
			{(1+\frac{4}{3}|\mathbf{x}|^2)^2}
	-\frac{\mu(\mathbf{x})}{\sqrt{I}}.
\end{equation}
The three terms in the kinetic energy are
kinetic energy for the size change,
for the rotation and for the shape change respectively.
The equation of motion for $I$ yields
Lagrange-Jacobi identity,
$\ddot{I}=4E+2U$.
From this equation, we get the following ``Saari's relation'' \cite{SaariCollisions},
\begin{equation*}
\frac{d}{dt}\left(
	\frac{I^2}{2}
	\frac{\frac{4}{3}|\dot{\mathbf{x}}|^2}
			{(1+\frac{4}{3}|\mathbf{x}|^2)^2}
\right)
=\sqrt{I}\,\frac{d\mu}{dt}.
\end{equation*}
Using the `time' variable $s$ defined by
\begin{equation}
\label{defS}
\frac{ds}{dt}
=\frac{1}{2I}\left( 1+\frac{4}{3}|\mathbf{x}|^2\right),
\end{equation}
the Saari's relation is written as
\begin{equation}
\label{Saari}
\frac{d}{ds}\left(\frac{1}{6}\left|\frac{d\mathbf{x}}{ds}\right|^2 \right)
=\sqrt{I}\,\frac{d\mu}{ds}.
\end{equation}
The equation of motion for $\mathbf{x}$ in terms of $s$ is
\begin{equation}
\label{eqOfMotion}
\frac{d^2 \mathbf{x}}{ds^2}
=\frac{\displaystyle 4C-\frac{8}{3}\mathbf{x}\wedge \frac{d\mathbf{x}}{ds}}
	{1+\frac{4}{3}|\mathbf{x}|^2}
	\left( \frac{dy}{ds}, -\frac{dx}{ds}\right)
	+ 3\sqrt{I}\,\frac{\partial \mu}{\partial \mathbf{x}}.
\end{equation}
Up to here,
we didn't assume $\mu=$ constant.

\section{Necessary condition}
\label{sec:nesCond}
Now, we consider a motion with $\mu=$ constant.
By the Saari's relation (\ref{Saari}), we have
\begin{equation}
\left|\frac{d\mathbf{x}}{ds}\right|
= v
\end{equation}
with constant $v \ge 0$.

For the case $v=0$, $d\mathbf{x}/ds=0$ then  $d^2\mathbf{x}/ds^2=0$.
The equation of motion (\ref{eqOfMotion}) yields $\partial \mu/\partial \mathbf{x}=0$.
Namely, the motion is homographic and the system stays one of the central configurations.

Let us examine the case $v>0$.
In this case, the point $\mathbf{x}(s)$ moves on the curve
$\mu(\mathbf{x})$ with finite speed $v$.
Since the number of points $\partial \mu/\partial \mathbf{x}=0$
are five, we can always take a finite arc 
on which $\partial \mu/\partial \mathbf{x} \ne 0$.
To keep satisfy $d\mu/ds=0$, the velocity $d\mathbf{x}/ds$
must be orthogonal to $\partial \mu/\partial \mathbf{x}$,
so we have
\begin{equation}
\label{velocity}
\frac{d\mathbf{x}}{ds}=\frac{\epsilon v}{|\partial \mu/\partial \mathbf{x}|}
	\left(-\frac{\partial \mu}{\partial y},
		\frac{\partial \mu}{\partial x}
	\right).
\end{equation}
Here, $\epsilon=\pm 1$ determines the direction of the motion.
Then, the acceleration (\ref{eqOfMotion}) is given by
\begin{equation}
\label{acceleration}
\frac{d^2\mathbf{x}}{ds^2}
=\Bigg(
	\frac{\epsilon v}{(1+4|\mathbf{x}|^2/3)|\partial \mu/\partial \mathbf{x}|}
	\left(
		4C
		-\frac{8\epsilon v}{3|\partial\mu/\partial \mathbf{x}|}
			\mathbf{x}\cdot\frac{\partial\mu}{\partial\mathbf{x}}
	\right)
	+3\sqrt{I}
\Bigg)
\frac{\partial\mu}{\partial\mathbf{x}}.
\end{equation}
Thus, the velocity (\ref{velocity}) and 
the acceleration (\ref{acceleration}) determine the curvature of this orbit
\begin{equation}
\kappa
=\frac{1}{1+4|\mathbf{x}|^2 /3}
	\left(
		-\frac{4C}{v}
		+\frac{8\epsilon}{3|\partial \mu/\partial \mathbf{x}|}
			\left(\mathbf{x}\cdot 
				\frac{\partial \mu}{\partial \mathbf{x}}
			\right)
	\right)
	-\frac{3\epsilon\sqrt{I}}{v^2}
		\left| \frac{\partial \mu}{\partial \mathbf{x}} \right|.
\end{equation}
On the other hand, 
the curve $\mu(\mathbf{x})=$ constant
has its own curvature,
\begin{equation}
\kappa
=\frac{\epsilon}{|\partial \mu/\partial \mathbf{x}|^3}
	\left(
		\left(\frac{\partial \mu}{\partial y}\right)^2
		\frac{\partial^2 \mu}{\partial x^2}
		-
		2\frac{\partial \mu}{\partial x}\frac{\partial \mu}{\partial y}
		\frac{\partial^2 \mu}{\partial x \partial y}
		+
		\left(\frac{\partial \mu}{\partial x}\right)^2
		\frac{\partial^2 \mu}{\partial y^2}
	\right).
\end{equation}
Equate the two expressions for $\kappa$, we have
a necessary condition for the motion,
\begin{eqnarray}
\fl
\sqrt{I}
=-\frac{4\epsilon C v}{3(1+4|\mathbf{x}|^2 /3)|\partial \mu/\partial \mathbf{x}|}
	+
	\frac{8v^2}{9(1+4|\mathbf{x}|^2 /3)|\partial \mu/\partial \mathbf{x}|^2}
		\left( \mathbf{x}\cdot \frac{\partial \mu}{\partial \mathbf{x}} \right)
	\nonumber\\
	-\frac{v^2}{3|\partial \mu/\partial \mathbf{x}|^4}
		\left(
		\left(\frac{\partial \mu}{\partial y}\right)^2
		\frac{\partial^2 \mu}{\partial x^2}
		-
		2\frac{\partial \mu}{\partial x}\frac{\partial \mu}{\partial y}
		\frac{\partial^2 \mu}{\partial x \partial y}
		+
		\left(\frac{\partial \mu}{\partial x}\right)^2
		\frac{\partial^2 \mu}{\partial y^2}
	\right).
\end{eqnarray}
This is the condition
that any motion with $\mu=$ constant and $d\mathbf{x}/dt \ne 0$
must satisfy.
The equation of motion is invariant under the scale transformation
$q_k \to \lambda q_k$ and $t \to \lambda^{3/2} t$.
This transformation makes $\sqrt{I} \to \lambda \sqrt{I}$,
$C \to \lambda^{1/2} C$,
$\mathbf{x} \to \mathbf{x}$,
$s \to \lambda^{-1/2}s$,
and $v \to \lambda^{1/2} v$.
Using this invariance, we can take $v=\sqrt{3}$ without loosing 
generality. We  write $C$ for $\epsilon C$.
Then, the necessary condition is
\begin{eqnarray}
\label{nesCondition}
\fl
\sqrt{I}
=-\frac{4C}{\sqrt{3}\,(1+4|\mathbf{x}|^2 /3)|\partial \mu/\partial \mathbf{x}|}
	+
	\frac{8}{3(1+4|\mathbf{x}|^2 /3)|\partial \mu/\partial \mathbf{x}|^2}
		\left( \mathbf{x}\cdot \frac{\partial \mu}{\partial \mathbf{x}} \right)
	\nonumber\\
	-\frac{1}{|\partial \mu/\partial \mathbf{x}|^4}
		\left(
		\left(\frac{\partial \mu}{\partial y}\right)^2
		\frac{\partial^2 \mu}{\partial x^2}
		-
		2\frac{\partial \mu}{\partial x}\frac{\partial \mu}{\partial y}
		\frac{\partial^2 \mu}{\partial x \partial y}
		+
		\left(\frac{\partial \mu}{\partial x}\right)^2
		\frac{\partial^2 \mu}{\partial y^2}
	\right),
\end{eqnarray}
and the energy is given by
\begin{equation}
\label{EbyI}
E=
\frac{1}{2}\left(\frac{d\sqrt{I}}{dt}\right)^2+\frac{C^2+1}{2I}-\frac{\mu}{\sqrt{I}}.
\end{equation}

Substituting  
$d\sqrt{I}/dt
	=(\partial \sqrt{I}/\partial \mathbf{x})
		\cdot(d\mathbf{x}/ds) (ds/dt)$, 
$d\mathbf{x}/ds$ in (\ref{velocity})
and the condition (\ref{nesCondition})
into this expression for the energy,
we will obtain the necessary condition 
expressed only by the shape variable $\mathbf{x}$.
However, the  condition (\ref{nesCondition}) in $\mathbf{x}$
turns out to be so complex to treat.
In the next section, we will rewrite the  condition (\ref{nesCondition})
in a concise form.

\section{Invariance of the necessary condition}
\label{sec:invariance}
Since we are considering equal mass case,
the theory is invariant under
the permutations of positions $\{q_i\}$.
The exchange of $q_1$ and $q_2$
makes $\zeta \to -\zeta$ and $\mathbf{x} \to -\mathbf{x}$.
The invariance of the necessary condition (\ref{nesCondition}) is
manifest.
On the other hand, the cyclic permutation 
$q_1 \to q_2 \to q_3 \to q_1$
makes
\begin{equation}
\label{theMap}
\zeta \to
\zeta'=\frac{3}{2}\, \frac{q_1}{q_3-q_2}
=\frac{1}{2}\frac{3/2+\zeta}{1/2-\zeta}.
\end{equation}
The invariance of (\ref{nesCondition}) under this transformation is not manifest.
In this section, we will rewrite the necessary condition
in a manifestly invariant form.

\subsection{Invariants}
%
Under the map (\ref{theMap}),
the Lagrange points $\zeta=\pm i \sqrt{3}/2$ are fixed
and the Euler points $\zeta=-3/2, 0, 3/2$ are cyclically permuted.
Let us define $\mu_k=I^{1/2}/r_{ij}$ for $(i,j,k)=(1,2,3)$, $(2,3,1)$ and $(3,1,2)$. 
Expressions by $\zeta$ are,
\begin{equation}
\label{def:muk}
\fl
\mu_1=\frac{1}{|\zeta-1/2|}\sqrt{\frac{1}{2}+\frac{2}{3}|\zeta|^2},\quad
\mu_2=\frac{1}{|\zeta+1/2|}\sqrt{\frac{1}{2}+\frac{2}{3}|\zeta|^2},\quad
\mu_3=\sqrt{\frac{1}{2}+\frac{2}{3}|\zeta|^2}.
\end{equation}
These three $\mu_k$ are also cyclically permuted by (\ref{theMap}). 
Note that the exchange $q_i \leftrightarrow q_j$
makes the exchange $\mu_i \leftrightarrow \mu_j$.
Therefore,
$\mu=\mu_1+\mu_2+\mu_3$ is invariant
under the permutations of $q_i$.

The kinetic energy for the shape change must be invariant.
Actually, we can easily check the invariance of
\begin{equation}
\label{metric1}
\frac{4}{3}\frac{|d\zeta|^2}{(1+\frac{4}{3}|\zeta|^2)^2}.
\end{equation}
So, it is natural to treat the space of $\zeta$
as a metric space whose distance is given by the equation (\ref{metric1}),
and the map (\ref{theMap}) is the isometric transformation.
Actually
Wu-Yi~Hsiang and Eldar~Straume \cite{Hsiang1995,Hsiang2006},
Alain~Chenciner and R.~Montgomery \cite{CMfigure8},
R.~Montgomery \cite{Montgomery2002},
and R.~Mockel \cite{MoeckelShooting}
showed that this metric space is
the ``shape sphere'' and 
the distance (\ref{metric1}) is the distance on the shape sphere.
Kenji~Hiro~Kuwabara and Kiyotaka~Tanikawa also noticed that
the shape sphere is useful
to investigate the equal-mass free-fall problem\cite{KuwabaraTanikawa,TanikawaKuwabara}.
The map (\ref{theMap}) makes the shape sphere
$2\pi/3$ rotation around the axis 
that connects the two Lagrange points.
The map $\zeta \to -\zeta$ makes $\pi$ rotation around the axis 
that connects one of the Euler point (corresponds to $\mathbf{x}=0$)
and one of two-body collision (corresponds to $\mathbf{x}=\infty$).

Let us use the notations in the tensor analysis.
We write $\zeta=x^1+ix^2$,
$\mathbf{x}=(x,y)=(x^1, x^2)$
and $\partial_i = \partial/\partial x^i$.
The metric tensor $g_{ij}$ and its inverse are
\begin{equation}
g_{ij}=\frac{4}{3}\frac{\delta_{ij}}{(1+\frac{4}{3}|\mathbf{x}|^2)^2},
\quad
\left(g_{ij}\right)^{-1}=g^{ij}=\frac{3}{4}\left(1+\frac{4}{3}|\mathbf{x}|^2\right)^2 \delta^{ij},
\end{equation}
where $\delta_{ij}=\delta^{ij}$ are the Kronecker's delta,
\begin{equation}
\delta_{ij}=\delta^{ij}=
\cases{
1	&for $i=j$,\\
0	&for $i\ne j$.
}
\end{equation}
Let $|g|$ be the determinant of $g_{ij}$,
\begin{equation}
|g|=\textrm{det}(g_{ij})=\frac{16}{9}\frac{1}{(1+\frac{4}{3}|\mathbf{x}|^2)^4}.
\end{equation}

As mentioned above, the configurational measure $\mu$ is invariant.
One obvious invariant is the magnitude of the gradient vector of $\mu$.
We write
\begin{equation}
|\nabla \mu|^2
=\sum_{i,j} g^{ij}(\partial_i \mu) (\partial_j \mu)
=\frac{3}{4}\left(1+\frac{4}{3}|\mathbf{x}|^2\right)^2
	\left| \frac{\partial \mu}{\partial \mathbf{x}} \right|^2.
\end{equation}
Therefore, the first term of the right hand side of
the necessary condition (\ref{nesCondition}) is
simply $-2C/|\nabla \mu|$.
The other obvious invariant is the Laplacian of $\mu$,
\begin{equation}
\Delta \mu
=\sum_{ij}\frac{1}{\sqrt{|g|}}\partial_i\left(g^{ij}\sqrt{|g|}\partial_j \mu\right)
=\frac{3}{4}\left(1+\frac{4}{3}|\mathbf{x}|^2\right)^2
	\frac{\partial}{\partial \mathbf{x}} \cdot \frac{\partial \mu}{\partial \mathbf{x}}
\end{equation}
Now, let us consider the following invariant,
\begin{equation}
\lambda=
\sum_{ij}g^{ij} (\partial_i \mu)(\partial_j |\nabla \mu|^2)
=\frac{3}{4}\left(1+\frac{4}{3}|\mathbf{x}|^2\right)^2
	\frac{\partial \mu}{\partial \mathbf{x}} \cdot \frac{\partial}{\partial \mathbf{x}}
	|\nabla\mu|^2.
\end{equation}
Explicitly performing the differentials, it yields
\begin{equation*}
\lambda
=3\left(1+\frac{4}{3}|\mathbf{x}|^2\right)^3
	\left(\mathbf{x}\cdot\frac{\partial \mu}{\partial \mathbf{x}}\right)
		\left|\frac{\partial \mu}{\partial \mathbf{x}}\right|^2
	+
	\frac{9}{16}
		\left(1+\frac{4}{3}|\mathbf{x}|^2\right)^4
    \frac{\partial \mu}{\partial \mathbf{x}} \cdot \frac{\partial}{\partial \mathbf{x}} 
    \left| \frac{\partial \mu}{\partial \mathbf{x}} \right|^2
\end{equation*}
Using this expression, the second and the third terms
in the necessary condition (\ref{nesCondition}) is simply
expressed as,
$\lambda/(2|\nabla \mu|^4) -\Delta \mu/|\nabla \mu|^2$.
Thus, the necessary condition is
expressed in the following invariant form,
\begin{equation}
\label{theNesCondInvariantForm}
\sqrt{I}
=-\frac{2C}{|\nabla \mu|}
	+\frac{\lambda}{2|\nabla \mu|^4}
	- \frac{\Delta \mu}{|\nabla \mu|^2}.
\end{equation}
The last obvious invariant what we will use is
\begin{equation}
\label{defD}
D\phi
=\frac{1}{\sqrt{|g|}}\sum_{i,j} \epsilon^{ij}
		(\partial_i \mu) (\partial_j \phi)
=\frac{3}{4}\left(1+\frac{4}{3}|\mathbf{x}|^2\right)^2
    \frac{\partial \mu}{\partial \mathbf{x}} \wedge \frac{\partial \phi}{\partial \mathbf{x}} 
\end{equation}
for any invariant $\phi$.
Where,
$\epsilon^{ij}$ is the Levi-Civita's anti-symmetric symbol, 
\begin{equation}
\epsilon^{ij}=
\cases{
1	&for $(i,j)=(1,2)$,\\
-1	&for $(i,j)=(2,1)$,\\
0	&for $i=j$.
}
\end{equation}
Then, using equations (\ref{defS}),
(\ref{velocity}) and (\ref{defD}), we have
\begin{equation}
\label{dandD}
\frac{d\phi}{dt}
=\frac{\epsilon}{I}\frac{D\phi}{|\nabla \mu|}.
\end{equation}

\subsection{Invariant variables}
For the Newton potential,
it is natural to use the variables
$r_1$ and $r_2$ defined by (\ref{def:r1andr2}).
Relations between $\mu_k$ defined in (\ref{def:muk}) and $r_1$, $r_2$
are
\begin{equation}
\label{eq:defOfMus}
\mu_1=r_1^{-1}\mu_3,\quad
\mu_2=r_2^{-1}\mu_3,\quad
\mu_3=\sqrt{(1+r_1^2+r_2^2)/3}.
\end{equation}
Now, consider the expression for the above invariants
$|\nabla\mu|^2$, $\Delta\mu$, $\lambda$
in terms of $r_1$ and $r_2$.
Let us write one of them $\psi(r_1, r_2)$.
It is composed by differentials of $\mu$ by $r_1$ or $r_2$
and products of $r_1$ and $r_2$.
Then, the result is composed of terms of
rational function of $\sqrt{(1+r_1^2+r_2^2)/3}$, $r_1$ and $r_2$,
namely $\mu_3$, $\mu_3/\mu_1$ and $\mu_3/\mu_2$.
Then, $\psi$ has the following form
\begin{equation}
\fl
\psi
=f(r_1,r_2)+g(r_1,r_2)\sqrt{\frac{1+r_1^2+r_2^2}{3}}
=f\left(\frac{\mu_3}{\mu_1},\frac{\mu_3}{\mu_2}\right)
+g\left(\frac{\mu_3}{\mu_1},\frac{\mu_3}{\mu_2}\right)\mu_3.
\end{equation}
Here, $f$ and $g$ represent some rational functions.
The function $\psi$ is invariant under the permutation of $q_i$,
namely the permutation of $\mu_i$.
So, it must be a ratio of some symmetric polynomials of $\mu_i$.
Therefore, it must have the following   expression
\begin{equation}
\label{invariantsInMuandRho}
\psi=h(\mu,\nu,\rho),
\end{equation}
where $h$ is a rational function of
elementary symmetric polynomials
\begin{equation}
\mu=\mu_1+\mu_2+\mu_3,\quad
\nu=\mu_1\mu_2+\mu_2\mu_3+\mu_3\mu_1,\quad
\rho=\mu_1\mu_2\mu_3.
\end{equation}
Expression in terms of $\mu_k$
or in terms of $\mu$, $\nu$, $\rho$ is not unique,
since, by the relation (\ref{eq:defOfMus}),
there is an identity
$\mu_1^{-2}+ \mu_2^{-2}+\mu_3^{-2}=3$.
Namely,
\begin{equation}
\label{identity}
\mu_1^2\mu_2^2+\mu_2^2\mu_3^2+\mu_3^2\mu_1^2
=3\mu_1^2\mu_2^2\mu_3^2.
\end{equation}
Therefore, we can eliminate $\nu$, using
\begin{equation}
\nu
=\sqrt{2\mu\rho+3\rho^2}.
\end{equation}
The expression of  $\psi=h(\mu,\sqrt{2\mu\rho+3\rho^2},\rho)$ is unique.
Thus,
the necessary condition will be expressed by
a function of  invariant shape variables $\mu$ and $\rho$.

Let us express $|\nabla \mu|^2$ by $\mu$ and $\rho$.
In terms of $r_i$, it is
\begin{equation}
\fl
|\nabla \mu|^2
=\frac{(1+r_1^2+r_2^2)^2}{3}
	\left(
		\left(\frac{\partial \mu}{\partial r_1}\right)^2
		+\left(\frac{\partial \mu}{\partial r_2}\right)^2
		+\frac{r_1^2+r_2^2-1}{r_1 r_2}
			\frac{\partial \mu}{\partial r_1}
			\frac{\partial \mu}{\partial r_2}
	\right).
\end{equation}
By a direct calculation, we get
\begin{eqnarray}
\fl
|\nabla \mu|^2
=\frac{1+r_1^2+r_2^2}{9r_1^4 r_2^4}
	\Bigg(
		2r_1^4r_2^4(r_1^2+r_2^2)\nonumber\\
		+r_1^4r_2^4(r_1+r_2)
		-r_1r_2(r_1^7+r_2^7)
		-r_1^4r_2^4
		-4r_1^3r_2^3(r_1+r_2)\nonumber\\
		+(2r_1^6+r_1^5r_2-r_1^4r_2^2
			-4r_1^3r_2^3
			-r_1^2r_2^4
			+r_1r_2^5
			+2r_2^6)\nonumber\\
		+r_1r_2(r_1^3+r_2^3)
		+2(r_1^4+r_2^4)
		-r_1r_2
	\Bigg).
\end{eqnarray}
Substituting  $r_1=\mu_3/\mu_1$ and $r_2=\mu_3/\mu_2$,
we obtain,
\begin{eqnarray}
\label{eq:nablaMuSquare1}
\fl
|\nabla \mu|^2
=\frac{(\mu_1^2\mu_2^2+\mu_2^2\mu_3^2+\mu_3^2\mu_1^2)}
	{9\mu_1^6 \mu_2^6 \mu_3^6}
	\Bigg(
		-(\mu_1^7\mu_2^7+\mu_2^7\mu_3^7+\mu_3^7\mu_1^7)
		\nonumber\\
		-\mu_1^4\mu_2^4\mu_3^4(\mu_1^2+\mu_2^2+\mu_3^2)
		-4\mu_1^4\mu_2^4\mu_3^4
			(\mu_1\mu_2+\mu_2\mu_3+\mu_3\mu_1)
		\nonumber\\
		+2(\mu_1^8\mu_2^4\mu_3^2+\dots)
		+(\mu_1^7\mu_2^4\mu_3^3+\dots)
	\Bigg).
\end{eqnarray}
In the last line,
dots in parentheses represent
similar 5 terms of  permutation
of $\mu_1, \mu_2, \mu_3$.
Then expressing by $\mu$, $\nu$, $\rho$,
we obtain a expression,
\begin{eqnarray}
\label{eq:nablaMuSquare2}
\fl
|\nabla \mu|^2
=\frac{\nu^2-2\mu\rho}{9\rho^6}
	\Bigg(
		-\nu ^7+7 \mu  \nu ^5 \rho +2 \mu ^4 \nu ^2 \rho ^2
		\nonumber\\
		-22 \mu ^2 \nu ^3 \rho ^2-3 \nu ^4 \rho ^2-4 \mu ^5 \rho ^3
		+24 \mu ^3 \nu  \rho ^3+18 \mu  \nu ^2 \rho ^3-27 \mu ^2 \rho ^4
	\Bigg).
\end{eqnarray}
As mentioned above,
the expressions (\ref{eq:nablaMuSquare1}) and (\ref{eq:nablaMuSquare2})
are not unique due to the identity (\ref{identity}).
Eliminating $\nu$, we finally get the 
following unique expression
\begin{equation}
\fl
|\nabla \mu|^2
=-\mu^2+2\mu^4+6\mu\rho-9\rho^2
	-3(2\mu^2-\mu\rho+3\rho^2)
		\sqrt{2\mu\rho+3\rho^2}.
\label{nablamu2}		
\end{equation}
Thus, we get the expression for $|\nabla \mu|^2$
in manifestly invariant variables $\mu$ and $\rho$.

By a similar way, $\Delta \mu$
in $(r_1, r_2)$ and $(\mu, \rho)$ are
\begin{equation*}
\fl
\Delta \mu
=\frac{(1+r_1^2+r_2^2)^2}{3}
	\left(
		\frac{1}{r_1}\frac{\partial}{\partial r_1}
			\left(r_1\frac{\partial\mu}{\partial r_1}\right)
		+
		\frac{1}{r_2}\frac{\partial}{\partial r_2}
			\left(r_2\frac{\partial\mu}{\partial r_2}\right)
		+
		\frac{r_1^2+r_2^2-1}{r_1 r_2} 
			\frac{\partial^2 \mu}{\partial r_1\partial r_2}
	\right),
\end{equation*}
\begin{equation}
\fl
\Delta \mu
=\mu+2\mu^3+6\rho
	-6\mu\sqrt{2\mu\rho+3\rho^2}.
\label{Deltamu}
\end{equation}
Similarly, the expressions for $\lambda$ are
\begin{equation*}
\fl
\lambda
=\frac{(1+r_1^2+r_2^2)^2}{3}
	\left(
		\frac{\partial\mu}{\partial r_1}\frac{\partial}{\partial r_1}
		+
		\frac{\partial\mu}{\partial r_2}\frac{\partial}{\partial r_2}
		+
		\frac{r_1^2+r_2^2-1}{2r_1 r_2}
		\left(
			\frac{\partial\mu}{\partial r_1}\frac{\partial}{\partial r_2}
			+
			\frac{\partial\mu}{\partial r_2}\frac{\partial}{\partial r_1}
		\right)
	\right)|\nabla \mu|^2,
\end{equation*}
\begin{eqnarray}
\fl
\lambda=\frac{1}{2}\Bigg(
4\mu^3-24\mu^5+32\mu^7-72\mu^2\rho+660\mu^4\rho+324\mu\rho^2
\nonumber\\
+36\mu^3\rho^2-432\rho^3+891\mu^2\rho^3+2349\mu\rho^4-243\rho^5
\nonumber\\
+3\Big(24\mu^3-60\mu^5-156\mu^2\rho+28\mu^4\rho+324\mu\rho^2
\nonumber\\
-93\mu^3\rho^2
-216\rho^3-27\mu^2\rho^3+81\mu\rho^4\Big)\sqrt{2\mu\rho+3\rho^2}
\Bigg).
\label{lambda}
\end{eqnarray}
Finally,
$(D\rho)^2$ is also invariant under the exchange of $q_i$,
therefore, it has an expression by $\mu$ and $\rho$,
\begin{equation*}
\fl
(D\rho)^2
=\frac{(1+r_1^2+r_2^2)^4 \Big(2(r_1^2+r_2^2)-(r_1^2-r_2^2)^2-1\Big)}{36r_1^2r_2^2}
	\left(
		\frac{\partial \mu}{\partial r_1}\frac{\partial \rho}{\partial r_2}
		-
		\frac{\partial \mu}{\partial r_2}\frac{\partial \rho}{\partial r_1}
	\right)^2,
\end{equation*}
\begin{eqnarray}
\fl
(D\rho)^2
=\frac{\rho^2(2\mu+3\rho)}{4}
	\Bigg(
		-(2\mu+3\rho)
		(4\mu^4+134\mu\rho-12\mu^3\rho-177\rho^2+9\mu^2\rho^2)
		\nonumber\\
		+2(28\mu^3+108\rho-36\mu^2\rho-45\mu\rho^2+54\rho^3)
			\sqrt{2\mu\rho+3\rho^2}
	\Bigg).
\end{eqnarray}

\section{Proof of the Saari's conjecture}
\label{sec:proof}
In the previous section, 
we find the expression for the necessary condition (\ref{theNesCondInvariantForm}) in terms of $\mu$ and $\rho$
by (\ref{nablamu2}), (\ref{Deltamu}) and (\ref{lambda}).
Since, we are assuming $\mu=$ constant,
time dependent variable is only $\rho$.
Therefore, $d \sqrt{I}/dt = (\partial \sqrt{I}/\partial \rho)(d\rho/dt)$.
Using (\ref{dandD}),
\begin{equation*}
\left(\frac{d \sqrt{I}}{dt}\right)^2
= \frac{1}{I^2} \frac{(D\rho)^2}{|\nabla \mu|^2}
			\left(\frac{\partial \sqrt{I}}{\partial \rho}\right)^2.
\end{equation*}
Substituting this expression and 
the necessary condition (\ref{theNesCondInvariantForm})
into the expression of the energy (\ref{EbyI}),
we obtain the  necessary condition for $\rho$ with three parameters $E$, $C$ and $\mu$,
\begin{equation}
\label{nesCond2}
E=\frac{1}{2I^2} \frac{(D\rho)^2}{|\nabla \mu|^2}
			\left(\frac{\partial \sqrt{I}}{\partial \rho}\right)^2
+\frac{C^2+1}{2I}-\frac{\mu}{\sqrt{I}}.
\end{equation}

If there is some finite motion with $\mu=$ constant and non-homographic,
this condition must be satisfied by some finite range of $\rho$.
However, since the right hand side of (\ref{nesCond2}) is analytic function of $\rho$,
the condition (\ref{nesCond2}) must be satisfied for all range of $\rho$.

In the vicinity of $\rho=0$, we have 
the expansion of (\ref{theNesCondInvariantForm})
\begin{equation}
\sqrt{I}=a_0+a_{1/2}\sqrt{\rho}+a_1 \rho + O(\rho^{3/2}),
\end{equation}
with
\begin{eqnarray*}
a_0=\frac{2(1-\mu^2+C\sqrt{-1+2\mu^2})}{\mu(1-2\mu^2)},\\
a_{1/2}=\frac{3\sqrt{2}\big((-2+\mu^2)\sqrt{-1+2\mu^2}-2C(-1+2\mu^2)\big)}
			{(1-2\mu^2)^2\sqrt{\mu(-1+2\mu^2)}},\\
a_1=\frac{3\big((-2+\mu^2)(1+6\mu^2)-2C(1+7\mu^2)\sqrt{-1+2\mu^2}\big)}
		{\mu^2(-1+2\mu^2)^3},
\end{eqnarray*}
and
\begin{eqnarray*}
|\nabla \mu|^2&=\mu^2(-1+2\mu^2)-6\sqrt{2}\mu^{5/2}+6\mu\rho+O(\rho^{3/2}),\\
(D\rho)^2&=-4\mu^6\rho^2+O(\rho^{5/2}).
\end{eqnarray*}
%
Then we obtain the power series expansion of (\ref{nesCond2}) 
by $\sqrt{\rho}$ up to the order $\rho$ at $\rho=0$.

The term of order $\rho^0$ in (\ref{nesCond2}) determine $E$.
Therefore, this order gives no information for $C$ and $\mu$.
%
The coefficient of order $\sqrt{\rho}$ is
\begin{equation}
0=\frac{-3\mu^{5/2}
		\Big(1+C\sqrt{-1+2\mu^2}\Big)^2
		\Big(
			(-2+\mu^2)-2C\sqrt{-1+2\mu^2}
		\Big)}
	{
		4\sqrt{2}
		\Big(
		-1+\mu^2-C\sqrt{-1+2\mu^2}
		\Big)^3
	}.
\end{equation}
The solutions $C$ of this equation are,
\begin{equation}
C=-\frac{1}{\sqrt{-1+2\mu^2}},\quad\frac{-2+\mu^2}{2\sqrt{-1+2\mu^2}}.
\end{equation}
%
For the case $C=-1/\sqrt{-1+2\mu^2}$,
\begin{equation}
\sqrt{I}
=\frac{2\mu}{-1+2\mu^2}
	+\frac{3\sqrt{2}\mu^{3/2}}{(-1+2\mu^2)^2}\sqrt{\rho}
	+\frac{9(1+2\mu^2)}{(-1+2\mu^2)^3}\rho
	+O(\rho^{3/2}),
\end{equation}
and the order $\rho^1$ coefficient in the  equation (\ref{nesCond2}) is
\begin{equation}
0
=-\frac{9\mu(-2+\mu^2)}{16(-1+2\mu^2)}.
\end{equation}
While the right hand side is always negative
since $\mu=\sqrt{(1+r_1^2+r_2^2)/3}\,(1+1/r_1+1/r_2)\ge 3$.
%
For the case $C=(-2+\mu^2)/(2\sqrt{-1+2\mu^2})$,
the coefficient $a_{1/2}$ vanish,
\begin{equation}
\sqrt{I}
=\frac{\mu}{4(-1+2\mu^2)}
	-\frac{3(-2+\mu^2)}{4(-1+2\mu^2)^3}\rho
	+O(\rho^{3/2}),
\end{equation}
and the coefficient of  order $\rho^1$ in the equation (\ref{nesCond2}) is
\begin{equation}
0=\frac{3\mu(-2+\mu^2)}{4(-1+2\mu^2)}.
\end{equation}
While the right hand side is always positive for $\mu\ge 3$.

Thus, there is no parameters $C$ and $\mu$
that satisfies the necessary condition (\ref{nesCond2}).
This completes the proof for the Saari's homographic conjecture.

\section{Discussions}
\label{sec:discussions}
We have proved the Saari's conjecture for equal-mass planar three-body problem
under the Newton gravity.

The symmetry under the permutation of the positions
$\{q_1, q_2, q_3\}$ has a  clucial role for our method.
For equal mass and  Newton potential case,
the necessary condition (\ref{theNesCondInvariantForm})   
is a symmetric rational function of $\mu_1$, $\mu_2$ and $\mu_3$.
Thus, it is a function of $\mu$ and $\rho$
as in equation (\ref{invariantsInMuandRho}).
This makes our proof simple.

The next step will be the case  with general mass ratio
and  general homogeneous potential 
$U=\sum m_i m_j /r_{ij}^\alpha$, $\alpha>0$.
For this case, 
however,
a invariant function under the permutation for suffix of bodies
will not have a simple form of 
manifestly invariant variables such as $\mu$ and $\rho$. 
We hope, someday, someone may find a proof for
the  conjecture for general mass ratio under the Newton potential
in some extension of our method.
On the other hand, we are afraid that
it is hard to extend our method to general $\alpha$.
We would have to find a completely new method for general $\alpha$.

\ack
This research of one of the author T.~Fujiwara  has been  supported by
Grand-in-Aid for Scientific Research 23540249 JSPS.

\end{document}